\documentclass[12pt]{article}

\usepackage{amssymb,latexsym}

\makeatletter \@addtoreset{equation}{section} \makeatother

\begin{document}

\begin{titlepage}

\thispagestyle{empty}

\begin{flushright}
\hfill{CERN-TH/2005-046} \\
\hfill{hep-th/0503122}
\end{flushright}

\vspace{35pt}

\begin{center}{ \LARGE{\bf 
Compactifications on twisted tori with fluxes \\[4mm]
and free differential algebras.
}}

\vspace{60pt}

{\bf  Gianguido Dall'Agata$^*$, \ Riccardo D'Auria$^{\#}$ \  and \ Sergio 
Ferrara$^{*\ddagger}$}

\vspace{15pt}

${}^{*}${\it  
Physics Department,\\
Theory Unit, CERN, \\
Geneva 23, CH1211,
Switzerland}

\vspace{15pt}

$^\#${\it Dipartimento di Fisica, Politecnico di Torino \\
C.so Duca degli Abruzzi, 24, I-10129 Torino, and\\
Istituto Nazionale di Fisica Nucleare, \\
Sezione di Torino,
Italy}

\vspace{15pt}

$^\ddagger${\it 
Istituto Nazionale di Fisica Nucleare, \\
Sezione di Frascati,
Italy\\
and\\
Department of Physics and Astronomy\\
University of California, Los Angeles, USA }

\vspace{50pt}

{ABSTRACT}

\end{center}

\vspace{20pt}

We describe free differential algebras for non--abelian one and two
form gauge potentials in four dimensions deriving the integrability
conditions for the corresponding curvatures.
We show that a realization of these algebras occurs in M--theory
compactifications on twisted tori with constant four--form flux, due
to the presence of antisymmetric tensor fields in the reduced theory.

\end{titlepage}

\newpage

\baselineskip 6 mm

\section{Introduction}

Flux compactifications on twisted tori provide interesting examples of
string and M---theory compactifications where most of the moduli
fields are stabilized
\cite{Scherk:1979zr,Kachru:2002sk,Schulz:2004ub,Kaloper:1999yr,Derendinger:2004jn,Antoniadis:2004pp,Bianchi:2005yz}.
Particular cases of such compactifications include heterotic string,
type II orientifold models and M--theory in the presence of constant
$p$--form fluxes (where $p$ depends on the particular string model and
$p=4$ in M--theory).
When fluxes and (or) Scherk--Schwarz geometrical fluxes are turned on,
interesting gauge algebraic structures emerge which in most cases have
the interpretation of a gauged Lie algebra 
\cite{Kaloper:1999yr,Andrianopoli:2002aq,DallAgata:2005ff,Andrianopoli:2005jv,Hull:2005hk}.

In this case the Maurer--Cartan equations (zero curvature conditions) 
read
\begin{equation}
d A^{\Lambda} + \frac12 \, f^{\Lambda}{}_{\Sigma\Gamma} A^{\Sigma} \wedge 
A^{\Gamma} = 0,
\label{MC}
\end{equation}
where integrability implies the Jacobi identities
\begin{equation}
f^{\Lambda}{}_{[\Sigma\Gamma}f^{\Pi}{}_{\Delta]\Lambda} = 0.
\label{Jacobi}
\end{equation}
This comes from the vanishing of the cubic term
\begin{equation}
d^2 A^{\Pi} = -\frac12\, f^{\Lambda}{}_{\Sigma\Gamma}f^{\Pi}{}_{\Delta\Lambda} A^{\Sigma} \wedge 
A^{\Gamma}\wedge A^{\Delta} = 0.
\label{howto}
\end{equation}

When fundamental tensor fields are present in the theory, in absence
of gauge couplings in the supergravity theory, one can transform them
into scalars and this is the way the full duality symmetry (sometimes
called U--duality) is recovered.
However, in presence of non--abelian gauge couplings, an obstruction
can arise in the dualization of such antisymmetric tensors, so that
the theory only preserves some subalgebra of the full duality group.
Moreover, the gauged algebra structure may be more complicated than an
ordinary Lie algebra and in fact, as noted in \cite{DallAgata:2005ff}
for generic Scherk--Schwarz and form flux couplings it turns out to be
a Free Differential Algebra (FDA)
\cite{sullivan,D'Auria:1982nx,Castellani:1991eu,vanNieuwenhuizen:1982zf,deAzcarraga:2002xi}.

In the case of M--theory, we will show that its Maurer--Cartan 
equations are equivalent to the integrability conditions for the 
4--form $G_{IJKL}$ and for the vielbein 1--form in D = 11.
This also will explain how the Lie algebra part of the Free 
Differential Algebra is deformed in the presence of generic 
Scherk--Schwarz and form flux couplings.

\section{The Free Differential Algebra and its Maurer--Cartan 
equations}

The generalization of (\ref{MC}) to a Free Differential Algebra
including 2--form gauge fields $B_{i}$ consists of the following
(zero--curvature) system
\begin{eqnarray}
{\cal F}^{\Lambda} &= & dA^{\Lambda} +\frac12\, f^{\Lambda}{}_{\Sigma\Gamma} A^{\Sigma} 
 \wedge A^{\Gamma} + m^{\Lambda i} B_i = 0, \label{dA}\\[2mm]
{\cal H}_i & = & dB_{i} + (T_{\Lambda})_{i}{}^{j} A^{\Lambda} \wedge B_{j} + 
 k_{i\,\Lambda \Sigma\Gamma} A^{\Lambda}\wedge A^{\Sigma} 
 \wedge A^{\Gamma}= 0, 
 \label{dB}
\end{eqnarray}
where $f^{\Lambda}{}_{\Sigma\Gamma}$, $(T_{\Lambda})_{i}{}^{j}$, $ 
m^{\Lambda i}$ and  $k_{i\,\Lambda \Sigma\Gamma}$ are the structure 
constants of the FDA.

The integrability condition of this system comes from the Bianchi 
identities
\begin{eqnarray}
d {\cal F}^{\Lambda} & = & 0, \label{dF}\\
d {\cal H}_{i} & = & 0. \label{dH}
\end{eqnarray}
From (\ref{dF}), by setting to zero the terms proportional to $A^{3}$ 
and $A \wedge B$ polynomials we get 
\begin{eqnarray}
&  & f^{\Lambda}{}_{\Sigma[\Gamma} f^{\Sigma}{}_{\Pi\Delta]} + 
2 m^{\Lambda i} k_{i\, \Gamma\Pi\Delta} = 0,\label{intda1}\\[2mm]
 &  &  f^{\Lambda}{}_{\Sigma\Gamma} m^{\Sigma j} + m^{\Lambda i} 
 (T_{\Gamma})_{i}{}^{j}=0,\label{intda2}
\end{eqnarray}
respectively.
From (\ref{dH}) we get three conditions from the vanishing of the terms
proportional to $B \wedge B$, $B \wedge A \wedge A$ and from $A^{4}$
terms:
\begin{eqnarray}
 &  & (T_{\Lambda})_{i}{}^{(j} m^{\Lambda k)} = 0, \label{intdb1}\\[2mm]
 &  & (T_{\Lambda})_{i}{}^{j} f^{\Lambda}{}_{\Sigma\Gamma} - 
2 (T_{[\Sigma})_{i}{}^{k}(T_{\Gamma]})_{k}{}^{j}+ 6 m^{\Lambda j} 
 k_{i\,\Lambda\Sigma\Gamma}=0,\label{intdb2}\\[2mm]
&& 3f^{\Lambda}{}_{[\Sigma\Gamma} k_{i\, \Pi\Delta]\Lambda} - 2
(T_{\Pi})_{i}{}^{j} k_{j\, \Sigma\Gamma\Delta]} = 0. \label{intdb3}
\end{eqnarray}

When $m^{\Lambda i} = 0$, the condition (\ref{intda1}) implies for 
the $A^{\Lambda}$ the ordinary Lie algebra Jacobi identities.
Equation (\ref{intdb2}) tells us that $(T_{\Lambda})_{i}{}^{j}$ is a 
representation of the Lie algebra and (\ref{intdb3}) states that 
$k_{i\Lambda\Sigma\Gamma}$ is a cocycle of the Lie algebra.
When $m^{\Lambda i} k_{i\Gamma\Pi\Delta} \neq 0$ (\ref{intda1}) gives 
the departure from an ordinary Lie algebra for the $f$ structure 
constants.

\section{FDA from M--theory on twisted tori with fluxes}

As an example of a concrete realization of the Free Differential
Algebra (\ref{dA})--(\ref{dB}), we will now describe the one obtained
by compactification of M--theory on twisted tori in the presence of
fluxes considered in \cite{DallAgata:2005ff}.
The compactification of M--theory to 4 dimensions provides 28 vector
fields $G_{\mu}^{I}$, $A_{\mu IJ}$ and 7 2--form tensor fields
$A_{\mu\nu I}$.
This means that we can identify the generic indices $\Lambda, i$ of
our FDA as follows $\Lambda = \{I, IJ\}$, $i = I$.
Furthermore one has to write the single indices $I, J$ in the same
position as $\Lambda, i$, but the antisymmetric couples $IJ$, $KL$, $
\ldots$ are written as upper indices if $\Lambda,\Sigma,\ldots$ are
lower ones and as lower indices if $\Lambda,\Sigma,\ldots$ are upper
ones.

If one considers first the case when only form fluxes are turned on, 
the Lie algebra is
\begin{equation}
\begin{array}{rcl}
[Z_I, Z_J] &=& g_{IJKL} W^{KL},\\[2mm]
[Z_I,W^{JK}] &=&  [W^{IJ},W^{KL}] = 0,
\end{array}
\label{Malgebraflux}
\end{equation}
which is the central extension of an abelian gauge algebra.
In this case the only non--vanishing structure constants are
\cite{DallAgata:2005ff}
\begin{eqnarray}
f^{\Lambda}{}_{\Sigma\Gamma} & = & f_{[IJ]\,KL} = g_{IJKL}\,, 
\label{fstruc1}\\
k_{i\Lambda\Sigma\Gamma} & = & k_{IJKL} =\frac16\, g_{IJKL}, \label{fstruc2}
\end{eqnarray}
while $m^{\Lambda i} = (T_{\Lambda})_{i}{}^{j} = 0$.
It then follows that (\ref{intda1}) and (\ref{intda2}) are trivially 
satisfied and $g_{IJKL}$ is arbitrary.
This result is a consequence of the very degenerate structure of the 
Lie algebra (\ref{Malgebraflux}).

An intermediate richer example comes in the case of Scherk--Schwarz 
fluxes $\tau_{IJ}^{K}$ and vanishing 4--form flux.
This is the case considered in the pioneering papers of 
Scherk--Schwarz \cite{Scherk:1978ta,Scherk:1979zr}.
In this case $k_{i\Lambda\Sigma\Gamma} = 0$, but $m^{\Lambda i}$ and 
$(T_{\Lambda})_{i}{}^{j}$ do not vanish.
In fact, the non--vanishing parts of these structure constants are
\begin{eqnarray}
m^{\Lambda i} \neq 0 & \hbox{ for } \Lambda = [IJ], \quad i = K, & 
m_{IJ}{}^{K} = \tau_{IJ}^{K}, \label{mSS}\\
(T_{\Lambda})_{i}{}^{j}\neq 0  & \hbox{ for } \Lambda= I, \quad i= J, 
\quad j = K,  & (T_{I})_{J}{}^{K} = -\tau_{IJ}^{K}. \label{TSS}
\end{eqnarray}
The other non--vanishing structure constants occur for 
$f^{\Lambda}{}_{\Sigma\Gamma}$ when
\begin{equation}
\begin{array}{rcl}
\Lambda = I, \Sigma = J, \Gamma = K &  & f^{I}{}_{JK} =\tau^{I}_{JK} 
\\[2mm]
\Lambda = [IJ], \Sigma = K, \Gamma = [LM] &  & f_{[IJ]\,K}{}^{[LM]} = 
- 2 \tau^{[L}_{K[I}\delta_{J]}^{M]}.
\end{array}
\label{fSS}
\end{equation}
In this case (\ref{intdb1}) is identically satisfied and 
(\ref{intda2}), (\ref{intdb2}) are identical to (\ref{intda1}), which 
reads as $\tau_{[IJ}^{L}\tau_{K]L}^{M} = 0$.
Note that $m^{\Lambda i}$ corresponds to a ``magnetic''' mass term 
for the $B_i$ field.

The $f^{\Lambda}{}_{\Sigma\Gamma}$ structure constants in (\ref{fSS}) 
define the Scherk--Schwarz algebra for M--theory:
\begin{equation}
\begin{array}{rcl}
[W^{IJ},W^{KL}] &=& 0, \\[2mm]
[Z_I, Z_J] &=& \tau_{IJ}^K Z_K,\\[2mm]
[Z_I,W^{JK}] &=&  2\tau^{[J}_{IL} W^{K]L}.
\end{array}
\label{MalgebraSS}
\end{equation}

Let us now consider the general case when both $\tau_{IJ}^{K}$ and 
$g_{IJKL}$ are non--vanishing.
In this case the last term in (\ref{intda1}) is non--vanishing for 
$\Lambda = [IJ]$, $\Sigma = K$, $\Gamma = L$ and $\Pi = M$.
It reads 
\begin{equation}
\tau_{IJ}^{N} g_{KLMN}.
\label{extra}
\end{equation}
If this term does not vanish the $f$ structure constants do not 
define a Lie algebra.
In this case (\ref{intda1}) (as also (\ref{intdb3})) becomes
\begin{equation}
\tau_{[IJ}^{N}g_{KLM]N} = 0.
\label{integrag}
\end{equation}
This condition has the 11--dimensional interpretation of the
integrability condition of the 4--form field strength
\cite{DallAgata:2005ff}.

All other equations are satisfied as a consequence of the $\tau$ 
Jacobi identities $\tau_{[IJ}^{L}\tau_{K]L}^{M} = 0$.
These follow from (\ref{intda1}) by taking $\Lambda,\Sigma,\Gamma,\Pi 
= IJKL$.
It is obvious that if the stronger condition (\ref{extra}) holds then 
the $f^{\Lambda}{}_{\Sigma\Gamma}$ define an ordinary Lie algebra.
This happens if the Scherk--Schwarz fluxes $\tau_{IJ}^{K}$ have the 
$K$ index complementary to the flux coupling $g_{IJKL}$.
This can actually be realized in certain type II orientifold models.

To summarize, we have shown that for generic Scherk--Schwarz 
couplings $\tau_{IJ}^{K}$ and 4--form flux $g_{IJKL}$, the M--theory 
gauge algebra is a Free Differential Algebra rather than an ordinary 
Lie algebra.
The equations
\begin{eqnarray}
\tau_{[IJ}^{M}\tau_{K]M}^{L} & = & 0, \label{tautau}\\
\tau_{[IJ}^{N}g_{KLM]N} & = & 0, \label{taug}
\end{eqnarray}
are the integrability conditions for the FDA.
When the stronger condition $\tau_{IJ}^{N} g_{KLMN} = 0$ holds then 
the $f^{\Lambda}{}_{\Sigma\Gamma}$ define an ordinary Lie algebra 
whose commutators read \cite{DallAgata:2005ff}
\begin{equation}
\begin{array}{rcl}
[Z_I, Z_J] &=& g_{IJKL} W^{KL}+ \tau_{IJ}^K Z_K,\\[2mm]
[Z_I,W^{JK}] &=&  2\tau^{[J}_{IL} W^{K]L},\\[2mm]
[W^{IJ},W^{KL}] &=&0.
\end{array}
\label{algebraM}
\end{equation}
It is interesting to note that in M--theory compactified on a twisted 
torus with 4--form flux turned on $m^{\Lambda i}$ and $g_{PQRS}$ have 
the physical interpretation of magnetic and electric masses for the 
antisymmetric tensors $B_{I}$.
This is clear looking at the covariant field strength
\begin{equation}
{\cal F}^{\Lambda} = dA^{\Lambda} +\frac12\, f^{\Lambda}{}_{\Sigma\Gamma} 
A^{\Sigma}\wedge A^{\Gamma}+ m^{\Lambda I} B_{I}.
\label{Ffs}
\end{equation}
This expression appears quadratically in the (kinetic parto of the) 
lagrangian together with the coupling
\begin{equation}
g_{IJKL} B_{M} \wedge dA_{NP} \epsilon^{IJKLMNP},
\label{ffa}
\end{equation}
which comes from the 11--dimensional Chern--Simons term $F\wedge F
\wedge A$.
It is amusing to note that the consistency condition
\cite{DAuria:2004yi,D'Auria:2004sy} for electric and magnetic
contributions to the mass is in this case a consequence of
(\ref{integrag}).

The M--theory FDA also includes a 3--form gauge field $C$ which is a 
singlet.
The zero--curvature condition for this 3--form is 
\begin{equation}
dC + m^{ij} B_i \wedge B_{j} + m_{\Lambda\Sigma}^{i} A^{\Lambda} 
\wedge A^{\Sigma} \wedge B_{i} + t_{\Lambda} A^{\Lambda} \wedge C + 
k_{\Lambda\Sigma\Gamma\Delta} A^{\Lambda} \wedge  A^{\Sigma} \wedge 
A^{\Gamma} \wedge  A^{\Delta}= 0.
\label{dC}
\end{equation}
In the M--theory FDA, the only non vanishing terms are $k_{IJKL} \sim 
g_{IJKL}$ and $m^{I}_{JK} \sim \tau^{I}_{JK}$, with all the other 
components and $t_{\Lambda}$ and $m^{ij}$ vanishing.
In this case the Bianchi identity is trivially satisfied because a 
5--form in D = 4 identically vanishes.
However, the curvature of $C$ can be determined by demanding its full 
invariance under all gauge transformations.

\section{Non--zero curvature case}

The previous Maurer--Cartan equations (\ref{dA})--(\ref{dB}), which 
entail the ``structure constants'' relations 
(\ref{intda1})--(\ref{intdb3}) can be lifted to non--zero curvature, 
so obtaining covariant Bianchi identities for the curvatures.
In the case of M--theory with Scherk--Schwarz fluxes turned on this 
procedure essentially reproduces the covariant curvatures $G$ of 
section 3.4 of \cite{Scherk:1979zr}.
When also the constant 4--form fluxes $F_{IJKL} = g_{IJKL}$ are 
turned on, then one gets generalized curvatures which are covariant 
under the combined 1--form and 2--form gauge transformations 
considered in section 2 of \cite{DallAgata:2005ff}.

An interesting new feature of the curvatures is the presence in ${\cal
H}_{I}$ of a ``contractible generator'' \cite{sullivan}, i.e.
in physical language, of a curvature itself (which also exists in the
ungauged theory)
\begin{equation}
{\cal H}_{I} = dB_I + {\cal F}^{J} \wedge A_{IJ},
\label{Hdef}
\end{equation}
where ${\cal F}^{J} = d A^{J}$.
This is a kind of Green--Schwarz (mixed) Chern--Simons term which 
modifies the gauge transformations of $B_{I}$ so that ${\cal H}_{I}$ is invariant 
under the gauge transformations
\begin{equation}
\delta B_{I} = d \Lambda_I - \epsilon_{IJ} {\cal F}^{J}, \quad \delta 
A^{I} = d \omega^{I}, \quad \delta A_{IJ} = d \epsilon_{IJ}.
\label{invgauge}
\end{equation}
The (ungauged) Bianchi identity is now
\begin{equation}
d {\cal H}_{I} = {\cal F}^{J} \wedge {\cal F}_{IJ},
\label{HIB}
\end{equation}
which satisfies $d^2 {\cal H}_I = 0$ and is also invariant under the gauge 
transformations (\ref{invgauge}).

Let us now consider the case when $g_{IJKL} \neq 0$ (but 
$\tau_{IJ}^{K} = 0$), so that 
\begin{equation}
{\cal F}_{IJ} = dA_{IJ} +\frac12\, g_{IJKL} A^{K} \wedge  A^{L}, \quad {\cal 
F}^{I} = dA^{I}.
\label{boh}
\end{equation}
Then the ${\cal H}_{I}$ curvature reads
\begin{equation}
{\cal H}_{I} = dB_{I} + {\cal F}^{J} \wedge A_{IJ} +\frac16\, g_{IJKL} A^{J}
\wedge A^{K} \wedge A^{L},
\label{Hflux}
\end{equation}
and the coefficient of the ${\cal F} \wedge A$ term is fixed, relative to 
the $A^{3}$ term in such a way that $d{\cal H}_{I} = {\cal F}^{J} \wedge 
{\cal F}_{IJ}$.
Now ${\cal H}_{I}$ and its Bianchi identity are invariant under the 
gauge transformations
\begin{eqnarray}
\delta B_{I} &=&d \Lambda_{I} - \epsilon_{IJ} {\cal F}^{J}+ \frac12\,
\omega^{M}g_{MIJK} A^{J} \wedge A^{K}, 
\label{deltaB}\\[2mm]
\delta A^{I} & = & d\omega^{I},  \\[2mm]
\delta A_{IJ} & = & d\epsilon_{IJ} - g_{IJKL} \omega^{K} A^{L}. 
\label{trasAA}
\end{eqnarray}

Analogously, the threefold antisymmetric tensor $C$ curvature is
\begin{equation}
dC - {\cal F}^{I} \wedge B_I + \frac{1}{4!}\,g_{IJKL} A^{I} \wedge
A^{J} \wedge A^{K} \wedge A^{L},
\label{dC}
\end{equation}
which is invariant under the gauge transformations
\begin{eqnarray}
\delta C & = & d\Sigma + {\cal F}^{I} \wedge \Lambda_I -\frac16\, g_{IJKL}
\omega^{I} \wedge A^{J} \wedge A^{K} \wedge A^{L}, \label{delC}\\
\delta B_{I} &=& d \Lambda_{I} -\epsilon_{IJ} {\cal F}^{J}+ \frac12\,
\omega^{M}g_{MIJK} A^{J} \wedge A^{K},  \label{delBI}\\
\delta A^I &  = & d\omega^I. \label{delai}
\end{eqnarray}
Note that the $dC$ field strength is a Lagrange multiplier and can be 
algebraically eliminated from the lagrangian giving a contribution to 
the scalar potential.

\section{Concluding remarks}

In the present paper we have considered the Free Differential Algebra 
which comes from M--theory compactified on a twisted torus with 
constant 4--form fluxes.
This is just a special case of the Maurer--Cartan equations described 
by (\ref{dA})--(\ref{dB}).
A similar situation arises in type IIA theories since in this case 
charged antisymmetric tensor fields are also present.
However, in this case one can find a particular set of geometrical 
fluxes which can be consistently set to vanish and then the Lie 
algebra structure is recovered because the condition 
\begin{equation}
m^{\Lambda i}k_{i \Sigma\Gamma\Pi} =0,
\label{mk}
\end{equation}
is satisfied.
Such examples were described in \cite{DallAgata:2005ff}.

The FDA given by the system of curvatures ${\cal F}^{\Lambda}$, ${\cal
H}_i$ can be recast in the form of an ordinary Lie algebra if (some of
the) $B_{i}$ are redefined so that the quadratic term in $A^{\Sigma}
\wedge A^{\Gamma}$ is absorbed in the new $\tilde B_i$ \cite{sullivan}.
This can be done at most for rank$(m)$ tensors fields, which can be
the same as the range of the $i$ indices provided that this is smaller
than that of the vector fields $\Lambda$, as in the M--theory case.
Explicitly, for those $B_{\alpha}$ for which the subblock
$m^{\alpha \beta}$ is invertible, one can introduce 
the definition ($\Lambda = \{\alpha,A\}$)
\begin{equation}
\tilde B_{\alpha}  \equiv  B_{\alpha} + \frac12\,
m^{-1}_{\alpha\beta} f^{\beta}{}_{\Lambda\Sigma} A^{\Lambda} \wedge A^{\Sigma},
\label{tildeB}
\end{equation}
so that the new zero curvature conditions read
\begin{eqnarray}
d\tilde B_{\alpha} &=& 0,\\[2mm]
{\cal F}^{\alpha} & = & d A^{\alpha} + m^{\alpha \beta} \tilde B_{\beta}=0,
\label{Ftilde}\\[2mm] 
{\cal F}^{A} & = & dA^{A} + \frac12
f^{A}{}_{BC} A^{B}\wedge A^{C}=0.
\label{Fnorm}
\end{eqnarray}
The new Lie algebra is defined by the structure constants 
$f^{A}{}_{BC}$ and this is obtained by deleting the 
$A^{\alpha}$ generators from the original algebra.
This is the quotient of the original algebra with the subalgebra 
related to the $A^{\alpha}$ vectors.
It is an obvious consequence of the Jacobi identities for ${\cal 
F}^{A}$ that $f^{A}{}_{\Lambda\Sigma} = 0$ whenever $\Lambda$ or 
$\Sigma$ take values in the $\alpha$ range.

In the M--theory case, the rank of $m^{\Lambda i}$ is encoded in the
Scherk--Schwarz fluxes $\tau_{IJ}^{K}$ regarded as a $7\times 21$
triangular matrix.
A quadratic submatrix can have at most rank 7 so the Lie algebra
spanned by the $A^{A}$ is at least $21$--dimensional.
When describing the algebra in terms of its generators, one must
delete the generators $W^{\tilde L \tilde K}$ whose gauge fields are
absorbed by the antisymmetric tensors.
The resulting Lie algebra is obtained by all $Z_{K}$, $W^{LK}$
generators but the $W^{\tilde L\tilde K}$, which is an Abelian subalgebra.
A simple example is the case when $\tau_{IJ}^{K}$ correspond to a
``flat group''.
In this case $B_{I} = \{B_{0}, B_{\alpha}\}$ and $A^{\Lambda} =
\{A^{\alpha}, A^{A}\}$, with $A^{\alpha} = A_{0\alpha}$ and 
$A^{A} = \{A^{0},A^{\alpha},A_{\alpha\beta}\}$.
The original structure constants follow from $\tau_{IJ}^{K} =
\tau_{0\beta}^{\alpha} =
{t^\alpha}_{\beta}$, where
$\alpha,\beta=1,\ldots,6$, and $t$ is an invertible
antisymmetric matrix, (this means that the 3 skew eigenvalues are
non--zero).
The redefined tensor fields are
\begin{eqnarray}
\tilde B_{\alpha} & = & {t^\delta}_{\gamma} \;t^{-1 
\beta}_{\alpha} \; A_{\beta\delta} \wedge A^{\tilde 
k} + A_{0\alpha} \wedge A^0 + B_{\alpha}, 
\label{tildeBtilde}\\
\tilde B_{0} & = & B_0 - A^{\alpha} \wedge A_{0\alpha}, 
\label{tildeb0}
\end{eqnarray}
and the zero curvatures conditions  read
\begin{eqnarray}
&&dA^{0}  =  0, \label{da0}\\[2mm]
&&dA^{\tilde\imath} + {t^\alpha}_{\beta} A^0 
\wedge A^{\beta}=0, \label{dai}\\[2mm]
 &  & dA_{0\alpha}+{t^\beta}_{\alpha}\tilde 
B_{\beta} = 0, \label{da0i}\\[2mm]
 &  & dA_{\alpha\beta} + 2 {t^\gamma}_{[\alpha} A^{0}\wedge A_{\beta]\gamma} = 0, \label{daij}\\[2mm]
 &  & d\tilde B_{\alpha} = 0, \label{dbi}\\[2mm]
 &  & d\tilde B_{0} - {t^\beta}_{\alpha} A^{\tilde 
\imath}\wedge A^{\gamma} \wedge A_{\beta\gamma} = 0. 
\label{db0}
\end{eqnarray}
Note that the Jacobi identities of the $\tau$ do not set any
constraint on the $t$ matrices.
If we split the generators into $Z_{0}$, $Z_{\alpha}$,
$W^{0\alpha}$, $W^{\alpha\beta}$, it is
immediate to see that the index $\alpha$ goes over six values
and the gauge fields $A_{\mu \,0\alpha}$ disappear from the
gauge algebra.
The generator algebra becomes then
\begin{equation}
\begin{array}{rcllrcl}
[Z_{0},W^{\alpha\beta}] &=& 2 \tau_{0\gamma}^{[\alpha}W^{\beta]\gamma}, && [Z_{0},Z_{\alpha}] &=&
\tau_{0\alpha}^{\gamma}Z_{\gamma},\\[4mm] [Z_{\alpha},
W^{\beta\gamma}] &=& [Z_{\alpha},Z_{\beta}] & =
&[W^{\alpha\beta},W^{\gamma\delta}]=0, &&
\end{array}
\label{algstr}
\end{equation}
which is the usual (22--dimensional) flat Scherk--Schwarz algebra.
This algebra becomes 24 or 26 dimensional if one or two eigenvalues of
the $t$ matrix vanish.
The same reasoning applies when form fluxes are present.
In this case the commutators of the $Z_{\alpha}$ are not
vanishing and the gauge algebra get modified.

Note that the physical interpretation of this reduction of the FDA to
a minimal part and a contractible one \cite{sullivan} corresponds to
the anti--Higgs mechanism where antisymmetric tensors absorb vector
fields to become (dual to) massive vectors.
The quotient Lie algebra is the unbroken gauge algebra.
It is interesting to see that, due to the cubic terms of the
1--forms $A^{\Lambda}$ in the ${\cal H}_{i}$ curvature (this only
happens when the 4--form flux is present), the quadratic part of the
${\cal F}^{\Lambda}$ curvature does not correspond to an ordinary Lie
algebra before the quotient has been taken.

Another interesting generalization is to extend such FDA to the 
fermionic sector of the theory, since the $D=4$ theory has $N=8$ 
local supersymmetry.
Such program was originally carried out in D=11 in \cite{D'Auria:1982nx} and 
its extension to the present compactification should be possible.

We finally remark that the different structures of the 4--dimensional
effective theories obtained when the gauge algebra is a FDA or an 
ordinary Lie algebra are reflected in different scalar potentials.
This fact may have important consequences when looking for complete 
moduli stabilization in such compactifications.


\bigskip \bigskip

\noindent
{\bf Acknowledgments}

\medskip

We would like to thank R.~Stora for an enlightening discussion.
The work of R.~D.~and  S.~F.~has been
supported in part by the European Community's Human Potential
Programme under contract HPRN--CT--2000--00131 Quantum Spacetime,
in which R.~D.~is associated to
Torino University and S.~F.~to INFN Frascati National Laboratories.
The work of S.F.~has also been supported in part by D.O.E.
grant DE-FG03-91ER40662, Task C.
S.~F. would like to thank the Department of Physics of the 
Polictecnico di Torino for the kind hospitality.


\providecommand{\href}[2]{#2}

\end{document}